\documentclass[%
 aip,
 jmp,%
 amsmath,amssymb,
 reprint,%
]{revtex4-1}

\usepackage{graphicx}% Include figure files
\usepackage{dcolumn}% Align table columns on decimal point
\usepackage{bm}% bold math
\begin{document}
%\preprint{AIP/123-QED}

\title{Hexagonal MASnI$_3$ exhibiting strong absorption of ultraviolet photons}

\author{Qiaoqiao Li}
\author{Wenhui Wan}
\author{Yanfeng Ge}
\author{Busheng Wang}
\author{Yingmei Li}
\author{Chuang Wang}
\affiliation{State Key Laboratory of Metastable Materials Science and Technology \& Key Laboratory for Microstructural Material Physics of Hebei Province, School of Science, Yanshan University, Qinhuangdao 066004, China }
\author{Yong-Hong Zhao}
\affiliation{College of Physics and Electronic Engineering, Center for Computational Sciences, Sichuan Normal University, Chengdu, 610068, China}
\author{Yong Liu}\email{yongliu@ysu.edu.cn, or ycliu@ysu.edu.cn}
\affiliation{State Key Laboratory of Metastable Materials Science and Technology \& Key Laboratory for Microstructural Material Physics of Hebei Province, School of Science, Yanshan University, Qinhuangdao 066004, China }

\begin{abstract}
  MASnI$_3$, an organometallic halide, has great potential in the field of lead-free perovskite solar cells. Ultraviolet photons have been shown to generate deep trapping electronic defects in mesoporous TiO$_2$-based perovskite, affecting its performance and stability. In this study, the structure, electronic properties, and optical properties of the cubic, tetragonal, and hexagonal phases of MASnI$_3$ were studied using first-principles calculations. The results indicate that the hexagonal phase of MASnI$_3$ possesses a larger indirect band gap and larger carrier effective mass along the \emph{c}-axis compared with the cubic and tetragonal phases. These findings were attributed to the enhanced electronic coupling and localization in the hexagonal phase. Moreover, the hexagonal phase exhibited high absorption of ultraviolet photons and high transmission of visible photons, particularly along the \emph{c}-axis. These characteristics demonstrate the potential of hexagonal MASnI$_3$ for application in multijunction perovskite tandem solar cells or as coatings in mesoporous TiO$_2$-based perovskite solar cells to enhance ultraviolet stability and photon utilization.
\end{abstract}

%\pacs{75.80.+q, 77.65.-j}

\maketitle
%\pacs{63.22.-m, 65.40.-b, 68.60.Dv, 63.20.kg}
% PACS, the Physics and Astronomy Classification Scheme.

%63.22.-m	Phonons or vibrational states in low-dimensional structures and nanoscale materials
%65.40.-b	Thermal properties of crystalline solids
%68.60.Dv	Thermal stability; thermal effects for thermal properties of thin films, see 68.60.Dv
%63.20.kg	Phonon-phonon interactions

\maketitle

Organometallic halide perovskites and Pb-based ones, in particular, show great potential as photon-harvesting materials in commercial solar cells since they were first reported in 2009~\cite{1}. Organometallic halide perovskite with the chemical formula ABX$_3$ (A = organic cation; B = metal; and X = halogen) exhibits a preeminent photovoltaic performance with a long carrier diffusion length, high charge carrier mobility~\cite{6,7,8}, superior point-defect properties~\cite{9,10}, and an ideal direct band gap that allows ABX$_3$ to utilize a wide range of photons~\cite{11,12}. Among the various organometallic halide perovskites, Pb-based perovskites have recently achieved high solar conversion efficiencies exceeding 20\%~\cite{2,3,4,5}. However, the need for toxic and poorly stable Pb$^{2+}$ to achieve such high performance hinders the commercialization of these perovskites~\cite{13,14,15,30,40}. In particular, mesoporous TiO$_2$-based perovskite is inherently unstable under ultraviolet (UV) illumination as a result of the deep trapping of injected electrons within the trap sites of TiO$_2$~\cite{40,34}. Post-transition metals with ns$^2$ electronic configurations beyond Pb$^{2+}$ (e.g., Sn$^{2+}$, Ge$^{2+}$, Sb$^{3+}$, and Bi$^{3+}$) are expected to be applied in lead-free perovskites in the future~\cite{30,43}.

The tolerance factor \emph{t} reflects the stability of the geometric crystal structure; \emph{t} should be between 0.813 and 1.107, with the ideal value being 1~\cite{16}. The \emph{t} of MAPbI$_3$ (approximately 0.84) is slightly smaller than that of MASnI$_3$ (0.85)~\cite{16}. Sn-based perovskites have been shown to possess better carrier mobilities (${10^2}$-${10^3}{cm^2}{V^{-1}}{S^{-1}}$) than Pb-based perovskites (${10}$-${10^2}{cm^2}{V^{-1}}{S^{-1}}$)~\cite{17,8}, which can be attributed to the delocalization of electronic states in Sn-based systems~\cite{19}. The maximum efficiency achieved by an Sn-based perovskite is 9.0\%~\cite{42}, much lower than that for Pb-based perovskites (22.1\%)~\cite{41}. This poor efficiency can be explained by the oxidation of Sn-based perovskite under ambient conditions or in the air. Zhang et al. reported that the formation energy of Sn-based perovskite is lower than that of Pb-based perovskite; in theory, this means that Sn-based perovskite has higher stability~\cite{41}. Xiang et al. recently proposed a new organic metal super-halide perovskite, MASnI$_2$BH$_4$, with a high electron affinity that reduces the likelihood of oxidation and enhances the stability in the presence of moisture~\cite{39}. Therefore, Sn-based perovskites show promise for the development of lead-free perovskites.

Arashdeep predicted a new hexagonal phase of MAPbI$_3$ (space group \emph{P6$_3$mc}) with a theoretically high dynamic stability~\cite{20}. Hexagonal FAPbI$_3$ was also observed experimentally at room temperature~\cite{17}. However, the hexagonal phase of MASnI$_3$ has not been studied in detail. In this study, we investigated the structure, electronic properties, and optical properties of the cubic (\emph{C}-phase), tetragonal (\emph{T}-phase) and hexagonal (\emph{H}-phase) phases of MASnI$_3$ using first-principles calculations. The results indicate that the \emph{H}-phase has a relatively small bond angle, resulting in a larger band gap and thus stronger absorption of UV photons compared to the \emph{C}- and \emph{T}-phases. Multijunction perovskite tandem is a kind of promising optical absorption layer applied in solar cells to optimize the bandgap for high efficiency~\cite{35,36}. \emph{H}-phase MASnI$_3$ can be used as an auxiliary absorption layer in tandem solar cells or as a coating for mesoporous TiO$_2$-based perovskite to improve the UV stability and enhance photon utilization.

Density functional theory calculations were performed using the Vienna ab initio simulation package~\cite{21} with the Perdew-Burke-Ernzerhof (PBE)~\cite{22} generalized gradient approximation as the exchange-correlation functional. Interactions between core and valence electrons were described by the projected augmented-wave pseudopotential~\cite{26} with an energy cutoff of 520 eV for the plane-wave basis. To describe van der Waals interactions, which are considered to be important in organometallic perovskites~\cite{23,24}, Grimme's DFT-D3 empirical dispersion correction to the PBE (Vdw-D3) was applied when calculating the energy-volume curves~\cite{25}. Structural optimization was performed with Hellmann-Feynman forces on atoms less than 0.001 eV/\AA. Three-dimensional k-point meshes ($5\times5\times5$ , $4\times4\times3$ and $3\times3\times4$ for the cubic, tetragonal, and hexagonal phases, respectively) were generated using the Monkhorst-Pack scheme~\cite{27}. When calculating optical properties, the NBANDS parameter was tripled relative to the default value to accommodate empty conduction bands, and the k-point meshes were expanded to $8\times8\times8$ (\emph{C}-phase), $6\times6\times5$ (\emph{T}-phase), and $5\times5\times6$ (\emph{H}-phase) to ensure accurate results.

\begin{table}[h] %The best place to locate the table environment is directly after its first reference in text
\newcommand{\PreserveBackslash}[1]{\let\temp=\\#1\let\\=\temp}
\newcolumntype{C}[1]{>{\PreserveBackslash\centering}p{#1}}
\newcolumntype{L}[1]{>{\PreserveBackslash\raggedright}p{#1}}
\newcommand{\tabincell}[2]{\begin{tabular}{@{}#1@{}}#2\end{tabular}} %tabincell{c}{the first line \\ the next\\the next\\ last} 单元格分行1/2格式
\caption{ Calculated and experimental lattice parameters of the \emph{C}-, \emph{T}-, \emph{H}- and \emph{O}-phases of MASnI$_3$ }\label{table1}
\begin{ruledtabular}
\begin{tabular}{lcccc}
Structure & \tabincell{c}{Space\\group} & \multicolumn{3}{c}{\tabincell{c}{Lattice parameters\\(\AA,deg)}}\\
& & PBE & Vdw-D3 & Exp~\cite{28} \\

\hline
Cubic(\emph{C}) & P\emph{m}3\emph{m} & $a=6.36$ & $a=6.06$ & $a=6.24$\\
Tetragonal(\emph{T}) & I4/\emph{mcm} & $a=8.83$ & $a=8.71$ & $a=8.73$\\
& & $c=12.93$ & $c=12.24$ & $c=12.50$\\

Hexagonal(\emph{H}) & P6$_3$\emph{mc} & $a=8.86$ & $a=8.77$ & \\
& & $c=7.91$ & $c=6.60$ & \\

Orthorhombic(\emph{O}) & P\emph{nma} & $a=8.48$ & $a=8.08$ & \\
& & $b=9.15$ & $b=8.50$ & \\
& & $c=12.67$ & $c=12.24$ & \\

\end{tabular}
\end{ruledtabular}
\end{table}

X-ray structure analyses indicated that the phase of MASnI$_3$ was cubic (space group P\emph{m}3\emph{m}) at temperatures above 295 K and tetragonal (space group I4/\emph{mcm}) at temperatures between 140 and 295 K~\cite{28}. Experimental investigations have found that the orthogonal phase (\emph{O}-phase) appears at lower temperatures (~110 K) ~\cite{28,44}; however, the structural parameters of the \emph{O}-phase could not be determined because of the limited experimental conditions. We performed density functional theory calculations with the PBE and Vdw-D3 functionals to determine the lattice parameters of the four MASnI$_3$ phases. The calculated and experimental lattice parameters are shown in Table~\ref{table1}. The theoretical lattice parameters of the \emph{C}- and \emph{T}-phases determined using both functionals are within 3.5\% of the experimental values. Since no experimental values are available for the \emph{O}-phase, we compared our calculated parameters with other theoretical results~\cite{39} and found a deviation of less than 4.0\%.

\begin{figure}[htb]
  \centering
  \includegraphics[width=.4\textwidth]{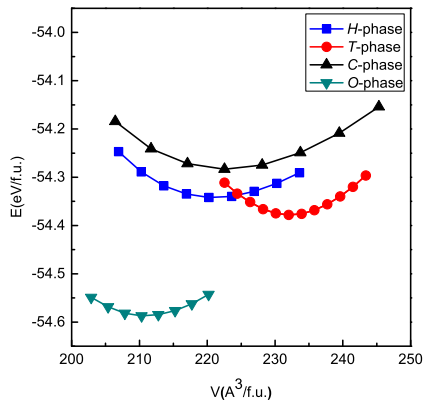}%
 \caption{Calculated total energy vs. volume per furmula unit (f.u.) of \emph{C}-phase, \emph{T}-phase, \emph{H}-phase and \emph{O}-phase MASnI$3$. Calculations used the Vdw-D3 functional.}\label{fig1}
\end{figure}

Vdw-D3 functional can effectively consider dispersion correction energy compared with PBE functional. Thus, to get a prccise energy comparison of the \emph{C}-, \emph{T}-, \emph{H}- and \emph{O}-phases, the volume-energy curves were calculated using the Vdw-D3 functional. Fig.~\ref{fig1} shows the calculated total energy per formula unit (f.u.) of the four phases as a function of unit cell volume. Notably, the trend in energy for the \emph{C}-, \emph{T}- and \emph{O}-phases of MASnI$_3$ is the same as that observed for MAPbI$_3$ (i.e., $\emph{C} > \emph{T} > \emph{O}$)~\cite{20,41} and follows the trend in transition temperature. However, the energy of the \emph{H}-phase falls between those of the \emph{C}- and \emph{T}-phases, indicating that the \emph{H}-phase may be metastable.

%\begin{figure}
%\includegraphics{E-V.eps} % Here is how to import EPS art
%\caption{\label{fig1} A figure caption. The figure captions are
%aut
%omatically numbered.}
%\end{figure}

\begin{figure}[htb]

  \centering
  \includegraphics[width=.45\textwidth]{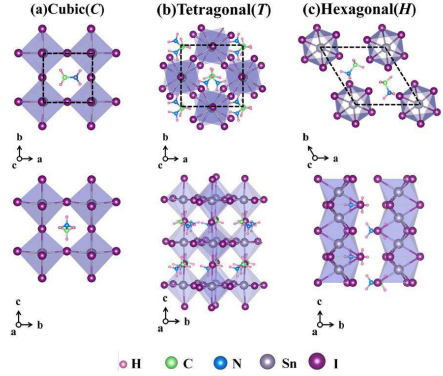}%[width=.7\linewidth]
 \caption{Polyhedral representations of the structures MASnI$_3$: (a)cubic phase (\emph{C}), (b) tetragonal phase (\emph{T}) and (c) hexagonal phase (\emph{H}). The unit cells are marked by the black dashed frames}\label{fig2}
\end{figure}

Considering that the \emph{O}-phase exists experimentally at extremely low temperatures, this phase has few practical applications. Thus, the remainder of this study focuses on comparing the characteristics of the \emph{C}-, \emph{T}-, and \emph{H}-phases using the PBE functional. Fig.~\ref{fig2} shows polyhedral representations of the \emph{C}-, \emph{T}-, and \emph{H}-phases of MASnI$_3$. Like conventional organometallic halide perovskite, the inorganic elements Sn and I constitute octahedral [SnI$_6^{4-}$] units; Sn cation is located at the center of the structure, and organic molecules are distributed outer these octahedral network gaps~\cite{29,30}. The \emph{C}- and \emph{T}-phases exhibit corner-connected and point-shared octahedra in which each I atom is shared by two Sn atoms [Fig.~\ref{fig2}(a) and (b)]. In contrast, the inorganic framework of the \emph{H}-phase [Fig.~\ref{fig2}(c)] comprises a combination of corner-connected, face-shared octahedra [SnI$_6^{4-}$], which consist of infinite octahedral [SnI$_6^{4-}$] chains along the \emph{c}-axis. This special connection is attributed to the smaller Sn-I-Sn bond angle in the \emph{H}-phase compared to in the \emph{C}- and \emph{T}-phases; the average Sn-I-Sn bond angles in the \emph{C}-, \emph{T}-, and \emph{H}-phases are $171.79^{\circ}$, $162.28^{\circ}$, and $75.63^{\circ}$, respectively.

\begin{figure}[htb]
  \centering
  \includegraphics[width=.5\textwidth]{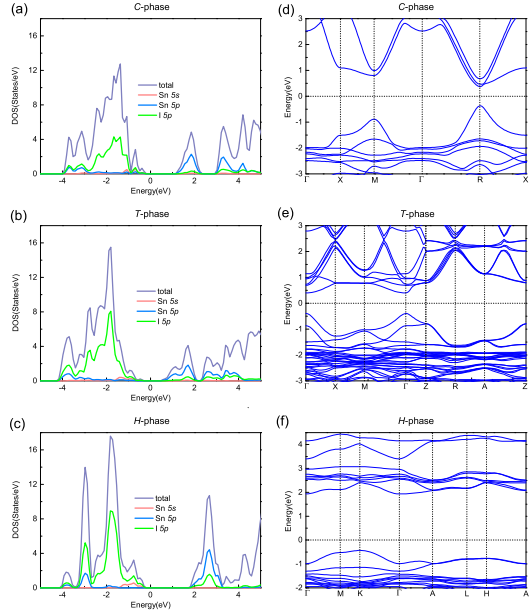}
  \caption{Total and partial DOSs (a)-(c) and band structures (d)-(f) of the \emph{C}-, \emph{T}-, and \emph{H}-phases of MASnI$_3$}\label{fig3}
\end{figure}

The density of states (DOS) and band structure provide important insights into the electronic properties of perovskites. Since the partial DOSs of C, N and H atoms corresponding their energy are quite low, they are barely involved in the formation of band edges. Fig.~\ref{fig3}(a)-(c) show the total and partial DOSs on metal and halide atoms in MASnI$_3$ calculated using the PBE functional. In all three cases, the main components of the band edges are semblable. The conduction band is dominated by the Sn-\emph{5p} and I-\emph{5p} electronic states. Meanwhile, the largest contribution to the valence band comes from the I-\emph{5p} electrons, while the Sn-\emph{5s} electrons make a small contribution. Fig.~\ref{fig3}(d)-(f) show band structure of each phase of MASnI$_3$ calculated using the PBE functional. The \emph{C}- and \emph{T}-phases both exhibit direct band gaps at the R and G points. In contrast, the \emph{H}-phase of MASnI$_3$ shows an indirect band gap indicated by the separate conduction band minimum and valence band maximum at the G and K points, respectively.

Table~\ref{table2} lists the calculated band gap value and type for each phase of MASnI$_3$. The GGA-PBE functional typically underestimates the band gaps of Sn-based perovskites as a result of spin-orbit coupling and many-body effects~\cite{8,47}. While band gap of the \emph{C}-phase calculated in this study (0.77 eV) is smaller than the experimental value (1.20 eV)~\cite{17}, it agrees with the theoretical value reported by Wang et al. (0.78 eV)~\cite{33}. The bandgap values followed the trend $\emph{C} < \emph{T} < \emph{H}$, with the band gap of \emph{H}-phase being much larger than those of \emph{C}- and \emph{T}-phases. Since the inorganic atoms of each phase have the same coordination number, this result can be attributed to the much smaller Sn-I-Sn bond angle of the \emph{H}-phase compared to the other two phases. This smaller bond angle results in stronger coupling between the I-\emph{5p}, Sn-\emph{5s} and Sn-\emph{5p} electron wave functions, and the larger band gap.

\begin{table}[h]
\begin{ruledtabular}
\caption{Bandgap values and types for the of the \emph{C}-, \emph{T}-, and \emph{H}-phases of MASnI$_3$.}\label{table2}
\begin{tabular}{lcc}
Structure & Band gap(eV) & Type\\
\hline
\emph{C}-phase & 0.77 & Direct\\
\emph{T}-phase & 0.79 & Direct\\
\emph{H}-phase & 2.36 & Indirect\\
\end{tabular}
\end{ruledtabular}
\end{table}
To further investigate the electronic properties of MASnI$_3$, the carrier effective masses of the \emph{C}-, \emph{T}-, and \emph{H}-phases were calculated using Eq.~\ref{equ1} and listed in Table~\ref{table3}:

\begin{eqnarray}
\frac{1}{m_i^*}&=&\frac{1}{\hbar^2}\frac{\partial^2E(k)}{\partial k_i^2} \quad (i=x,y,z)\label{equ1}
\end{eqnarray}

\begin{table}[h]
\newcommand{\PreserveBackslash}[1]{\let\temp=\\#1\let\\=\temp}
\newcolumntype{C}[1]{>{\PreserveBackslash\centering}p{#1}}
\newcolumntype{L}[1]{>{\PreserveBackslash\raggedright}p{#1}}
\begin{ruledtabular}
\caption{Carrier effective masses calculated for the \emph{C}-, \emph{T}-, and \emph{H}-phases of MASnI$_3$. $m_e^*$ is the effective mass for electron, $m_h^*$ is the effective mass for hole.}\label{table3}
\begin{tabular}{L{2.2cm}C{1.1cm}C{1.1cm}C{0.2cm}C{1.1cm}C{1.1cm}}
%\hline\hline
Structure & \multicolumn{2}{c}{$m_e^*/m_0$} &  & \multicolumn{2}{c}{$m_h^*/m_0$}\\
\cline{2-3}\cline{5-6}
& \emph{a}  &  \emph{c}  & &  \emph{a}  & \emph{c} \\
\hline
%\emph{C}-phase & \multicolumn{2}{c}{0.10} &  & \multicolumn{2}{c}{-0.18}\\
\emph{C}-phase & 0.60 & 0.10 & & -0.11 & -0.18\\
\emph{T}-phase & 1.58 & 1.07 & & -0.14 & -1.85\\
\emph{H}-phase & 0.45 & 3.09 & & -0.91 & -2.51\\
\end{tabular}
\end{ruledtabular}
\end{table}

Considering the anisotropic structures of the \emph{T}- and \emph{H}-phases, the effective masses along the \emph{a}- and \emph{c}-axes were calculated. The results suggest that the \emph{C}-phase has better carrier transport properties than the other phases as a result of its smaller carrier masses. The carriers in the \emph{H}-phase are more likely to propagate along the \emph{a}-axis because of the much smaller masses along the \emph{c}-axis. The effective hole mass of the \emph{T}-phase along the \emph{a}-axis is small (m$_h$ = -0.14 m$_o$), potentially indicating that the carriers are inclined to spread along this direction. The carrier effective masses of the \emph{H}-phase along the \emph{c}-axis are much larger, and this is still not difficult to analyze from the structure and DOS. The \emph{c}-axis is the face-shared direction of the [SnI$_6^{4-}$] octahedron; thus, the coupling and localization of electrons between Sn and I are more prominent than in the \emph{a}-axis direction of the \emph{H}-phase along with in the other point-shared phases. Therefore, the carrier effective masses are much larger in this direction.

\begin{eqnarray}
\alpha(\omega)&=&\frac{\sqrt{2}\omega}{c}[\sqrt{\varepsilon_1(\omega)^2+\varepsilon_2(\omega)^2}-\varepsilon_1(\omega)]^\frac{1}{2}\label{equ2}
\end{eqnarray}

\begin{eqnarray}
R(\omega)&=&\left|\frac{\sqrt{\varepsilon(\omega)}-1}{\sqrt{\varepsilon(\omega)}+1}\right|^2\label{equ3}
\end{eqnarray}

\begin{eqnarray}
L(\omega)&=&\frac{\varepsilon_2(\omega)}{\varepsilon_1(\omega)^2+\varepsilon_2(\omega)^2}\label{equ4}
\end{eqnarray}

\begin{figure}[htb]
  \centering
  \includegraphics[width=.5\textwidth]{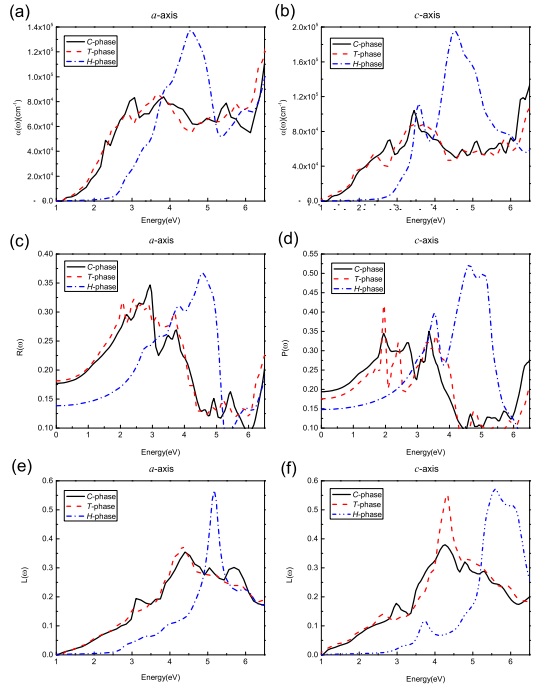}
  \caption{Calculated optical spectra of the \emph{C}-, \emph{T}-, and \emph{H}-phases of MASnI$_3$ in different directions: (a) and (b) the absorption coefficient
  $\alpha(\omega)$; (c) and (d) the reflectivity coefficient $R(\omega)$; and (e) and (f) the energy-loss coefficient $L(\omega)$.}\label{fig4}
\end{figure}

We also investigated the optical properties of \emph{C}-, \emph{T}-, and \emph{H}-phases of MASnI$_3$ by calculating the dielectric tensor $\varepsilon(\omega)$. the equation of $\varepsilon(\omega)$ is taken from Ref.~\cite{48}. Here, we use Eq.~\ref{equ2},~\ref{equ3},~\ref{equ4} to calculate the photon absorption coefficient $\alpha(\omega)$, reflection coefficient $R(\omega)$, and energy-loss coefficient $L(\omega)$ of these there phases, respectively. Also because of the anisotropy, the results along the \emph{a} and \emph{c} axis directions are shown in Figures.~\ref{fig4}.

%\begin{figure*}[htb]
%  \centering
%  \includegraphics[width=.7\textwidth]{optical}
%  %\includegraphics[width=.8\textwidth]{fanshe}\\
%  %\includegraphics[width=.8\textwidth]{sunshi}\\
%  \caption{Calculated optical spectra of the \emph{C}-, \emph{T}-, and \emph{H}-phases of MASnI$_3$ in different directions: (a) and (b) the absorption coefficient
%  $\alpha(\omega)$; (c) and (d) the reflectivity coefficient $R(\omega)$; and (e) and (f) the energy-loss coefficient $L(\omega)$.}\label{fig4}
%\end{figure*}

As shown in Fig. 4(a) and (b), photon absorption in the visible range (1.65-3.10 eV) is higher in the \emph{C}- and \emph{T}-phases than in the \emph{H}-phase as a result of the higher $\alpha(\omega)$ of the \emph{C}- and \emph{T}-phases in this range. However, the \emph{H}-phase shows more absorption and reflectivity than the \emph{C}- and \emph{T}-phases in the UV region [Fig.~\ref{fig4}(a)-(d)]. Particularly along the c-axis, $\alpha(\omega)$ is higher in the range of 3.5-6.0 eV [Fig.~\ref{fig4}(b)], and $R(\omega)$ is higher in the range of 3.4-5.7 eV [Fig.~\ref{fig4}(d)]. These results suggest that this \emph{H}-phase perovskite with high absorption and reflection could be applied in UV filter coatings~\cite{37} and used to improve UV stability. The low $\alpha(\omega)$ of the \emph{H}-phase in the visible region make it appropriate for use in multijunction perovskite tandem solar cells in combination with other perovskites that absorb well in the visible region. Fig.~\ref{fig4}(e) and (f) show that the $L(\omega)$ of the \emph{H}-phase is fairly low in the range of 1.65-3.10 eV, and its $R(\omega)$ [Fig.~\ref{fig4}(c) and (d)] is also significantly lower than those of the other two phases. These findings indicate that the \emph{H}-phase has favorable photon transmittance, which is beneficial for its application in multijunction perovskite tandem solar cells. Alternatively, the \emph{H}-phase of MASnI$_3$ could be simply used as a coating on mesoporous TiO$_2$-based perovskite layers to improve UV stability.

We studied the structural, electronic, and optical properties of the \emph{C}-, \emph{T}- and \emph{H}-phases of MASnI$_3$ using first-principles calculations. The results show that the \emph{H}-phase has a much smaller Sn-I-Sn bond angle than the \emph{C}- and \emph{T}-phases. The \emph{H}-phase also has face-shared octahedral [SnI$_6^{4-}$] chains aligned in the \emph{c}-axis direction. The calculated electronic properties of the three phases indicate that the band edges are formed primarily by the Sn-\emph{5s}, Sn-\emph{5p}, and I-\emph{5p} orbital electrons. The structural characteristics of the \emph{H}-phase result in enhanced electronic coupling and localization between Sn and I, resulting in a larger band gap and a larger carrier effective mass along the \emph{c}-axis compared with the \emph{C}- and \emph{T}-phases. However, the \emph{H}-phase exhibits high UV photon absorption and visible photon transmission, especially along the \emph{c}-axis. The \emph{H}-phase of MASnI$_3$ shows potential for application in multijunction perovskite tandem solar cell devices, or as a coating for mesoporous TiO$_2$-based perovskite layers to improve UV stability and photon utilization.

%\section{\label{sec:level1}RESULTS AND DISCUSSION}
%\subsection{\label{sec:level2}Crystal structure and lattice dynamics}

 This work was supported by the NSFC (Grants No.11747054 and No.11874273), the Specialized Research Fund for the Doctoral Program of Higher Education of China (Grant No.2018M631760), the Project of Heibei Educational Department, China (No. ZD2018015 and QN2018012), the Advanced Postdoctoral Programs of Hebei Province (No.B2017003004) and the Key Project of Sichuan Science and Technology Program (19SYXHZ0090).

%\indent\textcolor{blue}{\em Acknowledgments.}--


\nocite{*}
\bibliography{MASnI3-v1}
\end{document}